\documentclass[11pt, a4paper]{article}

\usepackage{amsmath}
\usepackage{amsfonts}
\usepackage{dsfont}
\usepackage{amssymb}
\usepackage{latexsym}
\usepackage{yhmath}
\usepackage{mathtools}
\usepackage{mathrsfs}
\usepackage{slashed}
\usepackage{cancel}
\usepackage[normalem]{ulem} % For IdentityMatrix
\usepackage{bm} % make argument bold
\usepackage{bbold}
\usepackage[]{breqn} % break equation
\usepackage{physics}
\usepackage{graphicx, rotating}
\usepackage{epstopdf}
\usepackage{epsfig}
\usepackage{graphics,subfigure}
\usepackage{comment}
\usepackage[utf8]{inputenc}
\usepackage{soul}
\usepackage{multirow}
\usepackage{color}
\usepackage[dvipsnames]{xcolor}
\usepackage{cite}
\usepackage{hyperref}

\hypersetup{colorlinks, citecolor=bluscuro, linkcolor=black, urlcolor=bluscuro}
\definecolor{rossos}{cmyk}{0,1,1,0.55}
\definecolor{bluscuro}{rgb}{0.15, 0.2, .85}
\definecolor{bluchiaro}{cmyk}{1,.3,0.,0.1}

\newcommand{\be}{\begin{equation}}
\newcommand{\ee}{\end{equation}}
\newcommand{\bea}{\begin{eqnarray}}
\newcommand{\eea}{\end{eqnarray}}

\newcommand\blfootnote[1]{%
  \begingroup
  \renewcommand\thefootnote{}\footnote{#1}%
  \addtocounter{footnote}{-1}%
  \endgroup
}

 \def\bea{\begin{eqnarray}}
  \def\eea{\end{eqnarray}}

	\def \beq {\begin{equation}}
	\def \eeq {\end{equation}}
	\def \ba {\begin{array}}
	\def \ea {\end{array}}

	\def \ecart {\noalign{\medskip}}
	\def \dis {\displaystyle}

\newcommand{\orcid}{\includegraphics{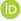}}
\newcommand{\orcidlink}[1]{\href{https://orcid.org/#1}{{\orcid}}}

\begin{document}

\begin{titlepage}
\begin{flushright}
{IFT-UAM/CSIC-24-144}
\end{flushright}
\begin{center} ~~\\
\vspace{0.5cm} 
\Large {\bf\Large
Nonlocality under Jaynes-Cummings evolution:\\
beyond pseudospin operators
\\[3mm]
}
\vspace*{1.5cm}

\normalsize{
{\bf 
Alexander Bernal
\orcidlink{0000-0003-3371-5320}
\blfootnote{
alexander.bernal@csic.es},
J.~Alberto Casas
\orcidlink{0000-0001-5538-1398}
\blfootnote{j.alberto.casas@gmail.com
} and
Jesús~M. Moreno
\orcidlink{0000-0002-2941-0690}
 } \\
 
\smallskip  \medskip
{\it Instituto de F\'\i sica Te\'orica, IFT-UAM/CSIC,}\\
\it{Universidad Aut\'onoma de Madrid, Cantoblanco, 28049 Madrid, Spain}}

\medskip

\vskip0.6in 

\end{center}

\centerline{ \large\bf Abstract }
\vspace{.5cm}

We re-visit the generation and evolution of (Bell) nonlocality in hybrid scenarios whose dynamics is determined by the Jaynes-Cummings Hamiltonian, a relevant example of which is the atom-cavity system. Previous approaches evaluate the nonlocality through the well-known qubit-qubit CHSH formulae, using combinations of pseudospin operators for the electromagnetic (EM) field observables. While such approach is sensible, it is far from optimal. In the present work we have used recent results on the optimal Bell violation in qubit-qudit systems, showing that the nonlocality is much greater than previously estimated, both with and without noise. We perform also an optimal treatment of the noise, so our results are optimal in this sense as well. We illustrate the results using different initial states for the EM field, including squeezed and coherent states. In addition, we study the asymptotic behavior of the entanglement. Remarkably, starting with a generic separable (pure) coherent state, the asymptotic (mixed) state is entangled, though does not violate Bell inequalities.

\vspace*{2mm}
\end{titlepage}

%%%%%%%%%%%%%%%%%%%%%%%%%%

\tableofcontents

\phantom{x}
\vspace{3cm}

\section{Introduction}
\label{sec:Intro}

Violation of Bell-like inequalities constitutes a paramount test of the non-classical character of nature, as they are incompatible with local hidden-variables theories. This violation has been experimentally confirmed in different setups, typically involving photons  \cite{Giustina:2015yza}, electrons \cite{Hensen:2015ccp} or superconducting circuits \cite{Storz:2023jjx}. On the other hand, systems exhibiting such Bell-statistics (a.k.a. Bell nonlocality)  also have application in developing quantum technologies, like device-independent quantum cryptography \cite{Zhang2022687,Nadlinger:2021uqf}, self-testing \cite{Supic20201}, evaluation of quantum resources \cite{Brunner:2013est,Chitambar:2018rnj}, etc. In this sense, modern quantum designs often involve hybrid systems, i.e. those made up of a discrete and a continuous (or, more generically, discrete and infinite) quantum subsystems.  A most representative example of hybrid systems is the one consisting of an atom, with two relevant energy levels, and an electromagnetic cavity, resonant with the two-level atom. 

There has been an increasing interest in exploring the (genuine) quantum characteristics of such atom-cavity systems, in particular entanglement \cite{PhysRevA.44.6023} and, more recently, Bell nonlocality \cite{Halder:2023iod}. More in particular, in the latter reference the authors consider an atom-cavity setup \cite{Blais:2020oqf}, evolving along the standard Jaynes-Cummings Hamiltonian \cite{Jaynes:1963zz,Shore19931195,Larson:2022jvs}, and explore the evolution of the CHSH inequality \cite{Clauser:1969ny} formulated in terms of certain pseudospin observables \cite{PhysRevLett.88.040406,ZHANG20111765}. This was done in the presence or absence of noise in the interaction and for several examples of states of the electromagnetic field, e.g. squeezed and coherent states.

In this paper we re-visit this setup, showing what are the optimal observables for Bell-violation (they are not pseudospin operators) and the optimal way to treat the noise. As a result, Bell nonlocality turns out to be much greater and more robust than previously estimated, even in the presence of noise. We illustrate this behavior with several types of initial states (including squeezed and coherent states).

In section \ref{sec:atom-cavity} we introduce the physical setup and the corresponding  evolution under the Jaynes-Cummings Hamiltonian. We also describe the different initial states considered in the following sections. In section \ref{sec:optimalBell} we review the optimal Bell inequalities for general qubit-qudit systems. In section \ref{sec:pure} we compute the Bell nonlocality for ideal pure states unitarily evolving according to the Jaynes-Cummings interaction. Section \ref{sec:noisy} is devoted to the treatment of noise. In particular we derive the corresponding evolution of nonlocality under noisy conditions. In section \ref{sec:asympt} we evaluate the asymptotic behaviour of nonlocality and entanglement for different types of initial states, including sqeezed and coherent states. In section \ref{sec:conclusions} we present the conclusions. Besides, we include two appendices, deriving some analytic relations for (pure) qubit-qudit states and the asymptotic behaviour of coherent states in the noisy scenario, respectively.

\section{The atom-cavity system}
\label{sec:atom-cavity}

Following the usual notation, let us denote by $\ket{g}, \ket{e}$ the ground and excited energy eigenstates of the atom subsystem, and by $\{\ket{n}\}_{n=0,1,2,\cdots}$ the Fock basis of the electromagnetic (EM) field in the cavity. The corresponding Hamiltonian has two pieces:
\bea
H=H_0+H_I\ .
\eea
$H_0$ is the free Hamiltonian, both for the atom and the electromagnetic field,
\bea
H_0=H_{\rm A}+H_{\rm EM} = \frac{1}{2}\hbar \omega\hat \sigma_z \otimes \mathbb{1}_{\rm EM} + \mathbb{1}_2\otimes \hbar\omega a^\dagger a,
\eea
where $\hbar \omega$ is the energy difference between the two atom levels, $\hat \sigma_z$ is the usual Pauli matrix operator in the atom basis, $\hat \sigma_z=\ketbra{e}-\ketbra{g}$ (other Pauli operators are analogously constructed), and $a, a^\dagger$ are the usual annihilation and creation operators of the EM field. Besides, $\mathbb{1}_2, \mathbb{1}_{\rm EM}$ are the identity operators in the atom and EM field Hilbert spaces respectively. Note that we have assumed that the cavity is resonant with the atom, thus $\omega$ is the same for both subsystems. On the other hand, the interaction Hamiltonian, $H_I$, is given by the usual Jaynes-Cummings expression:
\bea
H_I=\hbar\lambda(a+a^\dagger)(\hat \sigma_+ +\hat \sigma_-)
\label{H_I}, 
\eea
where $\lambda$ is a coupling constant which depends on the mode frequency, the modal volume of the cavity mode and the dipole moment of the atom\cite{Shore19931195,Larson:2022jvs}.
Working in the interaction picture, the states evolve along the effective Hamiltonian
\bea
H_{I, {\rm eff}}= e^{-iH_0 t/\hbar} H_I e^{iH_0 t/\hbar} \sim \hbar\lambda(a\hat \sigma_+ +a^\dagger\hat \sigma_-),
\eea
where we have neglected the fast-oscillating terms (i.e. as in the usual rotating wave approximation, RWA \cite{PhysRevLett.98.013601}). From now on we drop the label ``${\rm eff}$" from $H_I$.

The unitary evolution of the atom-cavity states from the above interaction Hamiltonian is straightforward (see, {\em e.g.} ref. \cite{PHOENIX1988381,gea1992new}) 
\begin{equation}
\begin{array}{rrcr}
U_\lambda(t)
\phantom{x}  = & 
e^{-iH_I t/\hbar}  =
e^{ -i \lambda t (a \hat \sigma_+ +  a^\dagger \hat \sigma_-)} &&  \\
= &
\cos \left( \sqrt{a a^\dagger  } \lambda t \right) 
      \ketbra{e}&-& 
i a \, 
\displaystyle {
                  \frac{\sin \left( \sqrt{ a^\dagger a } \lambda t \right)}
                       { \sqrt{ a^\dagger a } }
               }
        \ketbra{e}{g} \phantom{.} \\[3mm]
        & 
 - i a^\dagger\, 
 \displaystyle {
                  \frac{\sin \left( \sqrt{a a^\dagger } \lambda t \right)}
                       { \sqrt{a a^\dagger } }
               }
        \ketbra{g}{e}  &+&     \cos \left( \sqrt{ a^\dagger   a } \lambda t \right) 
      \ketbra{g}. 
\label{unitary evolution}
\end{array}
\end{equation}

Our main goal is to study the generation and evolution of entanglement and Bell nonlocality in the atom-cavity system, and the optimal way to experimentally test it. This requires to consider a generic initial state. For the pure case, the initial state has the form
\bea
\ket{\psi(0)} = C_g \ket{g}\ket{\varphi_g(0)}+C_e\ket{e}\ket{\varphi_e(0)},\ \ \ {\rm with} \ |C_g|^2+|C_e|^2=1,
\label{initial}
\eea
where $\ket{\varphi_g(0)},\ket{\varphi_e(0)}$ are two field states (not necessarily orthogonal). This is straightforwardly extended to the case of initial mixture states, described by a density matrix.

The field states in (\ref{initial}) have the generic form 
\bea
\ket{\varphi(0)} = \sum_{n=0}^\infty C_n\ket{n}, \ \ \ {\rm with} \ \ \sum_{n=0}^\infty |C_n|^2=1.
\eea
Following ref.\cite{Halder:2023iod} we will basically consider three types of initial field states:
\vspace{5mm}
\bea
\hspace{-6.2cm}\text{\em - Single-Fock states:}\ \ \ket{\varphi(0)} = \ket{k},\ \  {\rm i.e.}\hspace{1.8cm}C_n=\delta_{k,n},
\label{Fock}
\eea

\bea
\text{- \em Single-mode Squeezed  Vacuum (SMSV) states:}\ \ 
C_n=\left\{\begin{array}{cc}
 \displaystyle{
 (-1)^{\frac{n}{2}}\, \frac{\sqrt{n!}}{2^{\frac{n}{2}}\,\frac{n}{2}!}\frac{\left(e^{i\theta}\tanh (r)\right)^{\frac{n}{2}}}{\sqrt{\cosh (r)}}
 }
 & \text{$n$ even} \\[6mm]
	0  & \text{$n$ odd},
	\end{array}\right.
\label{SMSV}
\eea

\bea
\hspace{-4.5cm}\text{\em - Coherent states:}\hspace{5.5cm}
C_n=\frac{1}{\sqrt{n!}}e^{-\frac{|\alpha|^2}{2}}\alpha^n,
\label{Coherent}
\eea
\medskip

\noindent
where $k$, $\{r,\theta\}$ and $\alpha$ are (integer, real and complex) parameters.

\vspace{1mm}
In the absence of noise, the evolution of the generic initial state (\ref{initial}) is straightforwardly computed with the help of the unitary-evolution operator, (\ref{unitary evolution}),
\bea
\ket{\psi(t)} = U_\lambda(t)\ \ket{\psi(0)}.
\label{Upsi}
\eea
In general, even starting with a separable state this unitary evolution creates entanglement and non-locality between the two subsystems (the atom and the electromagnetic field). Actually, for pure states of any bipartite system (as the one at hand), it has been shown long ago \cite{Capasso:1973wt,Gisin:1991vpb} that entanglement implies non-locality. More specifically, given a pure entangled state, one can always tailor a Bell inequality (more precisely, a CHSH inequality, see next section) which is violated. This equivalence is lost when the state becomes a mixture, which is the usual case. Then, non-locality requires entanglement, but the reverse is not true. 

In any case, the experimental verification of the Bell violation shows that the system has non-local statistics at the same time as it certifies entanglement. For this purpose it is essential to design an optimal Bell test; otherwise, one is losing capacity to demonstrate these non-classical features.

\section{Optimal Bell (CHSH) inequalities for qubit-qudit}
\label{sec:optimalBell}

The most celebrated Bell test for non-locality is the violation of the CHSH inequalities \cite{Clauser:1969ny}: given a bipartite system, where Alice (Bob) can measure two observables, $A_1, A_2$ ($B_1, B_2$), which can take values $\{+1, -1\}$, the statistics of measurements is classical (i.e.
compatible with local realism) if and only if $|\langle {\cal O}_{\rm Bell}\rangle|\leq 2$ where 
\bea
{\cal O}_{\rm Bell}=A_1(B_1+B_2)+A_2(B_1-B_2).
\label{CHSH}
\eea

The CHSH inequalities are optimal for qubit-qubit systems (whether in a pure or a mixed state). Beyond that dimension, there exist other (typically more powerful) inequalities, like the CGLMP ones \cite{Collins:2002sun}, but the general question of optimally testing non-locality in arbitrary bipartite systems has not been solved. Fortunately, for qubit-qudit systems (which is essentially our case) it has been shown by Pironio \cite{Pironio_2014} that all the ``tight" Bell-like inequalities (those whose violation is a sufficient and necessary condition to violate local realism), involving two observables for both Alice and Bob, can be expressed as CHSH-type inequalities, 
\bea
|\langle {\cal O}_{\rm Bell}\rangle|\leq 2,
\eea
where ${\cal O}_{\rm Bell}$ is given by (\ref{CHSH}), and $A_1$, $A_2$ ($B_1$, $B_2$) are $2\times2$ ($d\times d$) Hermitian observables with eigenvalues $\{+1, -1\} $ ($\{+1, -1\} $ with some degeneracy). The problem arises because the space of Alice's and, especially, Bob's observables is enormous, and grows rapidly with increasing $d$ (Bob's dimension). Notice also that exploring that space entails to consider all the possible ``signatures" of $+1$s and $-1$s, which represents $d^2$ families of $B_1,B_2$ observables. Certainly, for the qubit-qubit system there is a well-known recipe by Horodecki et al. \cite{HORODECKI1995340} to evaluate the maximal violation and build the associated (optimal) observables $A_1$, $A_2$, $B_1$, $B_2$. For qubit-qudit systems (our case) the analogous prescription is the following \cite{Bernal:2024fio}.

First, express the density matrix of the system as
\begin{equation}
    \rho=\frac{1}{2}\left[\mathbb{1}_2\otimes\beta_0+\sigma_1\otimes\beta_1+\sigma_2\otimes\beta_2+\sigma_3\otimes\beta_3\right],
    \label{rho}
\end{equation}
where $\sigma_i$ are the standard Pauli matrices
and $\{\beta_0,\beta_i\}$ are $d\times d$ Hermitian matrices. The latter are easily obtained from $\beta_i  =  \Tr_A \left(\rho\, (\sigma_i \otimes \mathbb{1}_d   ) \right)$. Then, denoting by ${\cal R}$ an arbitrary $SO(3)$-rotation (characterized by three Euler angles), the maximum value of $\langle {\cal O}_{\rm Bell} \rangle$ is given by
\vspace{2mm}
\begin{equation}
\langle {\cal O}_{\rm Bell} \rangle_{\rm max}
= \dis{2\, \max_{{\cal R}} 
\left[
            \left(\sum_{i=1}^{d} |\lambda_i^{(1)}({\cal R})| \right) ^2  +
           \left( \sum_{i=1}^{d} |\lambda_i^{(2)}({\cal R})| \right) ^2
\right]^{1/2},} 
\label{generalresult}
\end{equation}
where $\lambda_i^{(1,2)} ({\cal R})$ stand for the eigenvalues of the first two $SO(3)$-rotated $\beta$ matrices, $({\cal R}\vec{\beta})_1, ({\cal R} \vec{\beta})_2$ with $\vec{\beta}=\left(\beta_1,\,\beta_2,\,\beta_3\right)$. The corresponding observables, $A_1$, $A_2$, $B_1$, $B_2$, are also straightforwardly obtained (for further details see ref. \cite{Bernal:2024fio}).

Consequently, in order to obtain the maximal Bell-violation it is enough to perform a scan in three angles in Eq.(\ref{generalresult}). Even without performing such a scan, i.e. by taking ${\cal R}=\mathbb{1}_3$ in (\ref{generalresult}), we obtain a lower bound in $\langle {\cal O}_{\rm Bell} \rangle_{\rm max}$ which is typically very close (at the 5\% level) to the actual maximum. In fact, the bound becomes even closer as the the dimension of Bob's increases, which is our case. 

The previous recipe, summarized in Eq. (\ref{generalresult}) is the one we have applied throughout the rest of paper.

\section{Evolution of nonlocality for  pure states}
\label{sec:pure}

\begin{figure}[htb]
\begin{center}
\begin{tabular}{ccc}
\includegraphics[width=8cm,clip=]{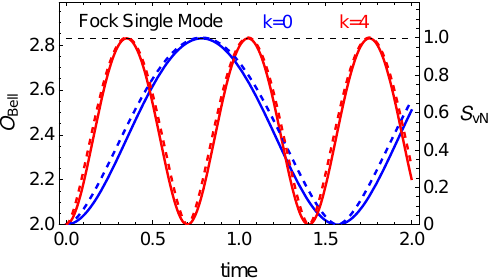}\\
\includegraphics[width=7.5cm,clip=]{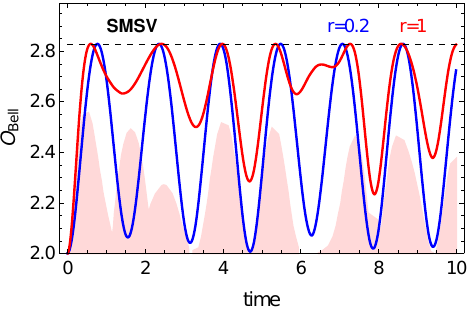} \hspace{0.4cm}\
\includegraphics[width=7.5cm,clip=]{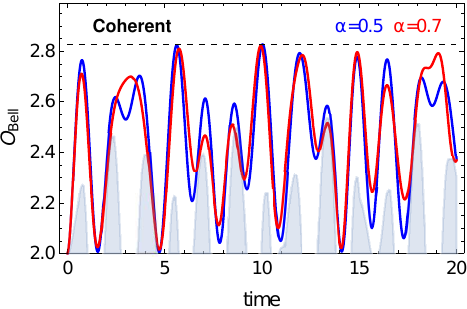}\hspace{0.4cm}
\end{tabular}
\caption{Upper side: 
$\langle {\cal O}_{\rm Bell} \rangle$ as a function of time for an initial product state made of the excited atom and a single cavity mode, $k$. We also display the von Neumann entropy (dashed lines), and check its close time-evolution (see text). Lower side: Evolution of  $\langle {\cal O}_{\rm Bell} \rangle$, as in upper panel side, but with a  Single-mode Squeezed Vacuum (left panel) or Coherent (right panel) initial cavity state. We also show  (shaded region) the results obtained in ref \cite{Halder:2023iod} in two particular cases (SMSV for $r=1$ and Coherent case for $\alpha =0.5$). }
\label{fig:pure}
\end{center}
\end{figure}

Although a realistic analysis of the atom-cavity system has to include the effect of noise, it is instructive to consider first the evolution of the states in the ideal case when the initial state is pure and the evolution is unitary according to the Jaynes-Cummings interaction, see Eqs.(\ref{H_I}, \ref{unitary evolution}, \ref{Upsi}). Suppose first that the initial state is separable, say
\bea
\ket{\psi(0)} = \ket{e}\ket{\varphi_e(0)},
\label{initial separable}
\eea
where $\ket{\varphi_e(0)}$ is one of the three possibilities described in Eqs.(\ref{Fock}, \ref{SMSV}, \ref{Coherent}). In order to simultaneously study the evolution of the entanglement and non-locality we will use the most standard measurement of entanglement for a pure state, namely the von Neumann entropy of the reduced density matrix,
\bea
S_{\text{vN}}=-\Tr (\rho_A \ln\rho_A),
\eea
where $\rho_A=\Tr_B \rho$ (the result is identical whether one traces in Bob's or Alice's indices). Besides, the non-locality is given by $\langle {\cal O}_{\rm Bell} \rangle_{\rm max}-2$, where $\langle {\cal O}_{\rm Bell} \rangle_{\rm max}$ is given by Eq.(\ref{generalresult}). For squeezed (SMSV) and coherent states the computation of both quantities requires to sum an infinite series; however, given the exponential suppression of large-$n$ contributions in (\ref{SMSV}, \ref{Coherent}), for practical (numerical) purposes in our case it is enough to consider $n_{\max}={\cal O}(10)$ in all the expressions. Fig. \ref{fig:pure} shows the evolution of $S_{\text{vN}}$ and $\langle {\cal O}_{\rm Bell} \rangle_{\rm max}$ in the three cases (single-Fock, squeezed and coherent) for different values of the parameters. As expected, entanglement and non-locality are generated along the unitary evolution, exhibiting a quasi-periodic pattern. The results are very similar if we start with a more complicated (but still pure) state, e.g. an entangled state. This is logical since, as mentioned, entanglement is generated anyway along the evolution.

It is interesting to note that for the case considered in this section (pure states) we can derive a fully analytic result for $\langle {\cal O}_{\rm Bell} \rangle_{\rm max}$. Expressing the state $\ket{\psi(t)}$ as a linear combination 
\bea
\ket{\psi(t)}=\ket{g}\ket{\varphi_g}+\ket{e}\ket{\varphi_e}
\label{decomp}
\eea
(which is trivial since $\ket{\varphi_{g,e}}=\bra{g,e}\ket{\psi(t)}$), after some algebra (for details see Appendix A) we arrive at the following result:
\bea
\langle {\cal O}_{\rm Bell} \rangle_{\rm max}=2\sqrt{1+4\left(
\bra{\varphi_g}\ket{\varphi_g}\bra{\varphi_e}\ket{\varphi_e}
-|\bra{\varphi_g}\ket{\varphi_e}|^2\right)}.
\label{exact}
\eea
Note that in general $\ket{\varphi_g},\ket{\varphi_e}$ are neither orthogonal nor normalized, but they satisfy 
$\bra{\varphi_g}\ket{\varphi_g}+\bra{\varphi_e}\ket{\varphi_e}=1$. Incidentally, when the initial EM field state is a Single-Fock or a squeezed state, the states $\ket{\varphi_g},\ket{\varphi_e}$ are actually orthogonal at any time, so the combination (\ref{decomp}) is actually a Schmidt decomposition.
The results obtained using the analytic expression (\ref{exact}) or the general formula (\ref{generalresult}) are of course identical.

We have also plotted in  Fig.~\ref{fig:linear} the von Neumann entropy $S_{\text{vN}}$ (for the Single-Fock case). Clearly $S_{\text{vN}}$ follows $\langle {\cal O}_{\rm Bell} \rangle_{\rm max}$ closely, and this behaviour is completely analogous for the other two cases. This can be analytically understood. As explained in Appendix A, for two-qubit systems there is a one-to-one relationship between $S_{\text{vN}}$ and $\langle {\cal O}_{\rm Bell} \rangle_{\rm max}$, which is plotted in Fig.~\ref{fig:linear}. Since the atom-cavity pure state can be expressed as a Schmidt decomposition at any time, this two-qubit relationship applies, thus the closeness of the $S_{\text{vN}}$ and $\langle {\cal O}_{\rm Bell} \rangle_{\rm max}$ curves.

To conclude this section, let us briefly compare the previous results with those obtained in ref. \cite{Halder:2023iod}. In that paper the non-locality issue was examined with a different strategy, which consisted of defining a set of pseudospin operators \cite{PhysRevLett.88.040406,ZHANG20111765}, $S^q_i$, where $i=1,2,3$ and $q$ is an (integer) fixed parameter. These operators act on (essentially) all states of the energy basis, $\{\ket{n}\}$ and obey the same algebra as the Pauli matrices.
Then the Bob's observables were constructed as linear combinations of these three operators (plus the identity), in a way similar to the qubit-qubit system. The advantage of this procedure is that one can straightforwardly apply the Horodeckis's result for CHSH-violation. This is completed by optimizing in the value of $q$. The limitation is that the method is far from optimal since it takes into account a very reduced set of possible observables. As discussed above, the optimal strategy is the one depicted in the previous paragraphs. To illustrate this point we have included in the Fig. \ref{fig:pure} a couple of (shaded) curves showing the results from \cite{Halder:2023iod}.

\section{Evolution with noise}
\label{sec:noisy}

Let us now be more realistic by taking into account that the real process in the lab is affected by noise and uncertainties of different types. Following ref. \cite{Halder:2023iod}, we will consider the uncertainties and fluctuations associated with the coupling $\lambda$, e.g. from defects in the cavity and/or non-uniformity of the EM field. However, our treatment can be straightforwardly extended to any other source of uncertainty or noise.

To be specific, let us suppose that the probability distribution of $\lambda$ is given by a normal distribution, $P(\lambda) = {\cal N}(\bar \lambda, \sigma)$, where $\bar \lambda$ is the mean value. Provided the typical time of fluctuation is much larger than the interval in which the measurements are performed, we can assume that the value of $\lambda$ is constant along the entire process, though affected by the above Gaussian uncertainty.

Now, in order to explore the nonlocality of the system, a rational way to proceed \cite{Halder:2023iod} is to operate as if the process were the ideal one ($\lambda=\bar\lambda$). This means that at every time $t$ we evaluate the ideal state $\ket{\psi(\bar\lambda, t)}$ and the corresponding optimal observables $A_1(\bar\lambda, t), A_2(\bar\lambda, t), B_1(\bar\lambda, t), B_2(\bar\lambda, t)$. In the ideal case the measurement of these observables (in a large number of shots) would give us the maximal value of the Bell operator, $\langle \psi(\bar\lambda, t)| {\cal O}_{\rm Bell}(\bar\lambda, t)|\psi(\bar\lambda, t)\rangle$, according to Eq.(\ref{exact}). In the real (noisy) case we still use in this approach the same $A_i(\bar\lambda, t), B_i(\bar\lambda, t)$ observables, thus getting a distribution:
\be
\langle  {\cal O}_{\rm Bell}(\bar\lambda, t)\rangle = \int\ d\lambda P(\lambda)\ \langle\psi(\lambda, t)| {\cal O}_{\rm Bell}(\bar\lambda, t)|\psi(\lambda, t)\rangle.
\ee
This was the procedure followed in ref.\cite{Halder:2023iod}, but actually there is a better (and optimal) approach. Due to the uncertainty in $\lambda$, the state of the system at any time can be described by a density matrix
\be
\rho(t)= \int\ d\lambda\ P(\lambda)\ |\psi(\lambda, t)\rangle\langle\psi(\lambda, t)|,
\label{DensMat}
\ee
which is straightforwardly computable for a given probability distribution. So, the total uncertainty about the state is encoded in the form of $\rho(t)$. Then we apply the recipe (\ref{generalresult}) to obtain the maximal value of $\langle  {\cal O}_{\rm Bell}\rangle$. The procedure is not equivalent to the previous one since the optimal observables are not $A_i(\bar\lambda, t), B_i(\bar\lambda, t)$. They can be however straightforwardly obtained from $\rho(t)$ \cite{Bernal:2024fio}.

\begin{figure}[htb]
\begin{center}
\begin{tabular}{ccc}
\includegraphics[width=7.5cm,clip=]{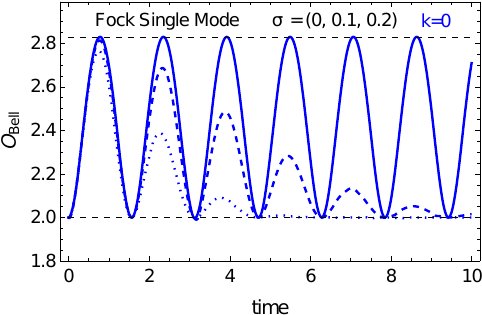} \\[4mm]
\includegraphics[width=7.5cm,clip=]{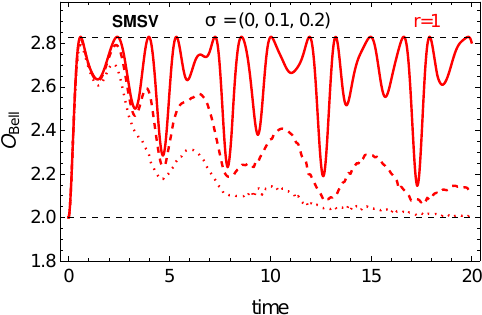} 
\includegraphics[width=7.5cm,clip=]{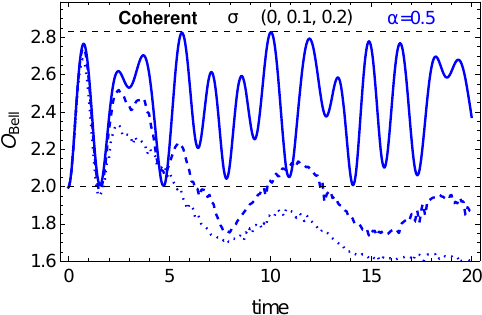} \hspace{0.1cm}
\\
\end{tabular}
\caption{The same as Fig.~\ref{fig:pure}, but including the noise due to the uncertainty in the 
Jaynes-Cummings coupling, $\lambda$. We assume a normal distribution, with mean    $\bar \lambda=1$ and three values of the variance:   $\sigma =0$ (solid curve), $\sigma =0.1$ (dashed) and  $\sigma =0.2$ (dotted). The first case corresponds to the absence of noise shown in Fig.~\ref{fig:pure}.}
\label{fig:noisy}
\end{center}
\end{figure}

We have shown in Fig.~\ref{fig:noisy} the evolution of $\langle  {\cal O}_{\rm Bell}\rangle_{\rm max}$, obtained in the latter way, for
different values of the variance of $P(\lambda)$, and for the three scenarios of Fig. \ref{fig:pure}, i.e.
when the  initial state is $\ket{\psi(0)} = \ket{e}\ket{\varphi_e(0)}$, with $\ket{\varphi_e(0)}$ a Single-Fock state, a squeezed state or a coherent state.
As expected, $\langle  {\cal O}_{\rm Bell}\rangle_{\rm max}$ decreases with time, but there is a temporal window to certify nonlocality and thus entanglement. Remarkably, for the Single-Fock and (especially) squeezed cases, nonlocality is preserved all along the way (except at some isolated times), although $\langle  {\cal O}_{\rm Bell}\rangle_{\rm max}$ approaches 2 (loss of nonlocality). Therefore these experimental setups are advantageous to test Bell violation.

In the specific case of the Single-Fock state, $\ket{\varphi_e(0)} = \ket{k}$,
$\rho(t)$ effectively reduces to a qubit-qubit density matrix and the value of $\langle  {\cal O}_{\rm Bell}\rangle_{\rm max}$ is easily computed in an analytical way, namely 
\bea
\langle {\cal O}_{\rm Bell}(t) \rangle_{\rm max}=2\sqrt{1+e^{-4 (k+1)\, t^2\, \sigma^2}\sin ^2\left(2\sqrt{k+1}\, t\, \bar \lambda\right)}.
\label{FockExact}
\eea
For fixed values of the parameters $(\bar \lambda,\sigma)$ it is clear that, except for a discrete set of values in time, Bell violation is always attained. Nevertheless, this violation is exponentially attenuated for long times, matching the tendency shown in Fig.~\ref{fig:noisy}.

\section{Asymptotic behaviour}
\label{sec:asympt}

In the previous section we analyzed the temporal evolution of $\langle  {\cal O}_{\rm Bell}\rangle_{\rm max}$, illustrating its behaviour for finite time in Fig.~\ref{fig:noisy}. A complementary approach is given in this section, in which we explore the asymptotics of the time-evolution for a  general 
pure initial state (either separable or entangled)
\bea
\ket{\psi(0)}=\ket{g}\ket{\varphi_g(0)}+\ket{e}\ket{\varphi_e(0)},
\eea
with the initial EM field states being 
\bea
\ket{\varphi_g(0)} &=& \sum_{n=0}^\infty C_{g,\,n}\ket{n}, \ \ \ {\rm with} \sum_{n=0}^\infty |C_{g,\,n}|^2=|C_g|^2,\nonumber \\ \ecart
\ket{\varphi_e(0)} &=& \sum_{n=0}^\infty C_{e,\,n}\ket{n}, \ \ \ {\rm with} \sum_{n=0}^\infty |C_{e,\,n}|^2=|C_e|^2
\label{init}
\eea
and $|C_g|^2+|C_e|^2=1$.  Applying the time evolution operator, we get
\bea
\ket{\varphi_g(\lambda,t)} &=& \sum_{n=0}^\infty\left[ C_{g,\,n}\cos\left(\sqrt{n}\,t\,\lambda \right)- i\, C_{e,\,n-1}\sin\left(\sqrt{n}\,t\,\lambda \right)\right]\ket{n},\\ \ecart
\ket{\varphi_e(\lambda,t)} &=&  \sum_{n=0}^\infty\left[ C_{e,\,n}\cos\left(\sqrt{n+1}\,t\,\lambda \right)- i\, C_{g,\,n+1}\sin\left(\sqrt{n+1}\,t\,\lambda \right)\right]\ket{n}.
\eea
The matrix density at any time shall, computed via Eq.\eqref{DensMat}, in general leads to convoluted expressions involving nontrivial products of sines and cosines.  However, since we are interested in the long-time regime most of these terms are asymptotically vanishing, notably simplifying the final expression.  In particular, upon the integration (\ref{DensMat}) only the $\cos^2$, $\sin^2$  terms survive in the $t\to\infty$ limit. In the end, the density matrix in the asymptotic regime is described by the block decomposition:
\be
\rho(t\to\infty)=\ketbra{e}\otimes \rho_{ee}+\ketbra{e}{g}\otimes \rho_{eg}+\ketbra{g}{e}\otimes \rho_{eg}^\dagger+\ketbra{g}\otimes \rho_{gg},
\label{DensMatInfty}
\ee
where
\bea
\rho_{ee}&=& \frac{1}{2}\sum_{n=0}^\infty |C_{e,\,n}|^2\,\ketbra{n}+\frac{1}{2}\sum_{n=0}^\infty |C_{g,\,n+1}|^2\,\ketbra{n},\\ \ecart
\rho_{eg}&=& \sum_{n=0}^\infty \Re{C_{g,\,n+1}\,C_{e,\,n}^\ast}\,\ketbra{n}{n+1}, \\ \ecart
\rho_{gg}&=& \frac{1}{2}|C_{g,\,0}|^2\,\ketbra{0}+\frac{1}{2}\sum_{n=0}^\infty |C_{g,\,n}|^2\,\ketbra{n}+\frac{1}{2}\sum_{n=0}^\infty |C_{e,\,n-1}|^2\,\ketbra{n}.
\eea
Note that, if the coefficients defining the initial state, $C_{g,\,n}, C_{e,\,n}$ are such that 
$\rho_{eg}$ vanishes, the resulting asymptotic density matrix is  separable, independently on the entanglement and nonlocality acquired for finite time. Let us now discuss in more detail the asymptotic limit for separable initial states, $C_{g,\,n}=C_g\, C_n$ and $C_{e,\,n}=C_e\,C_n$, with $C_n$ taking the values introduced in section \ref{sec:atom-cavity}, Eqs. \eqref{Fock}, \eqref{SMSV}, \eqref{Coherent}.

\subsection*{Single-Fock states}
Starting with a separable Single-Fock state, $C_n=\delta_{k,n}$, the asymptotic density matrix reads
\be
\begin{aligned}
\rho(t\to\infty)=\ &\frac{1}{2}\ketbra{e}\otimes\left(|C_{e}|^2\ketbra{k}+|C_{g}|^2(1-\delta_{k,0})\ketbra{k-1}\right)+\\ \ecart
&\frac{1}{2}\ketbra{g}\otimes\left(|C_{e}|^2\ketbra{k+1}+|C_{g}|^2(1+\delta_{k,0})\ketbra{k}\right).
\end{aligned}
\label{FockInfty}
\ee
This example matches the condition for asymptotic separability given in last subsection, $\rho_{eg}=0$,  even though the state is entangled and nonlocal at finite times, as it was shown in Eq.\eqref{FockExact}. Actually, this result is very general. The only initial state for which the asymptotic density matrix is entangled is given by 
$C_{g,\,n}=\delta_{k+1,n}\,C_g$ and $C_{e,\,n}=\delta_{k,n}\,C_e$, with both $C_g,\,C_e\neq0$ (which can be taken as real
after a basis redefinition). This corresponds to (a particular form of) an entangled initial state. Then the asymptotic density matrix becomes
\be
\rho(t\to\infty)=\frac{1}{2}\ketbra{e}\otimes \ketbra{k}+\frac{1}{2}\ketbra{g}\otimes \ketbra{k+1}+C_g\,C_e \left(\ketbra{e}{g}\otimes\ketbra{k}{k+1}+\ketbra{g}{e}\otimes\ketbra{k+1}{k}\right).
\label{FockInftyEnt}
\ee
A sufficient (and in this case also necessary) condition for entanglement is given by the Peres-Horodecki 
criterium  \cite{Peres:1996dw,Horodecki:1996nc}, i.e. the presence of negative eigenvalues in the partial transpose of $\rho(t\to\infty)$. Applying this criterium, it is easy to check that the above asymptotic density matrix is entangled
whenever $C_g\,C_e\neq0$.

\subsection*{SMSV states}
The asymptotic behaviour of SMSV states is similar to the Fock ones. Namely, starting with a separable state with $C_n$ as given by Eq.\eqref{SMSV}, it is straightforward to check that $\rho_{eg}$ vanishes, so the asymptotic state is separable (though for intermediate times is entangled and nonlocal).

\subsection*{Coherent states}
Contrary to the previous cases in which initial separable states evolve into asymptotic separable ones, coherent states exhibit a different behaviour. Recall that for a separable initial state, $C_{g,\,n}=C_g\, C_n$ and $C_{e,\,n}=C_e\,C_n$
in Eq.~(\ref{init}), where, for a coherent state,
\be
C_n=\frac{1}{\sqrt{n!}}e^{-\frac{|\alpha|^2}{2}}\alpha^n.
\ee
By redefining the $\{\ket{g},\ket{e}\}$ and $\{\ket{n}\}$ bases it is possible to take  $C_g,\,C_e$ and $\alpha$ real without loss of generality.  In order to check that $\rho(t\to\infty)$ is entangled we apply the Peres-Horodecki criterium.
Actually, taking the partial transposed of $\rho$, Eq.~(\ref{DensMatInfty}), with respect to the EM-field Hilbert space leads to
\be
\rho^{\mathrm{T}_B}(t\to\infty)=\ketbra{e}\otimes \rho_{ee}+\ketbra{e}{g}\otimes \rho_{eg}^{\mathrm{T}}+\ketbra{g}{e}\otimes \rho_{eg}+\ketbra{g}\otimes \rho_{gg}.
\ee
Independently on the value of $C_g,\,C_e$ and $\alpha$, this matrix always has a negative eigenvalue associated with a state of the form $\ket{\phi_n}=A\,\ket{g}\ket{n}+B\,\ket{e}\ket{n+1}$, with $|A|^2+|B|^2=1$ and $n=n\left(C_g,\,C_e,\,\alpha\right)$. The details of the computation of the eigenvalue are displayed in Appendix B. It is worth-mentioning that although the state is entangled, it does not violate the CHSH inequality, as the asymptotic $\langle {\cal O}_{\rm Bell} \rangle_{\rm max}$ value rapidly goes below 2 (Fig.~\ref{fig:noisy}).

\section{Conclusions}
\label{sec:conclusions}

We have re-visited the generation and evolution of entanglement and, especially, (Bell) nonlocality in hybrid scenarios whose dynamics is determined by a standard Jaynes-Cummings Hamiltonian, a relevant example of which is the atom-cavity system. For this task we have considered different types of initial states of the electromagnetic (EM) field (Fock, squeezed and coherent states) both in the absence and (more realistically) in the presence of noise. Previous approaches to this problem \cite{Halder:2023iod} evaluate the nonlocality through the well-known qubit-qubit CHSH formulae \cite{HORODECKI1995340}, using combinations of pseudospin operators for the CHSH EM-field observables. While such approach is sensible, it is far from optimal. In the present work we have used recent results on the optimal Bell violation in qubit-qudit systems \cite{Bernal:2024fio}, showing that the nonlocality is much greater than previously estimated, both with and without noise. 

On the other hand, for the treatment of noise we have included the possible uncertainties (in particular, the one in the value of the Jaynes-Cummings coupling) in the density matrix that describes the joint system. This guarantees that the evaluation of the nonlocality is optimal. Typically, the nonlocality of the system exhibits a quasi-periodic pattern for pure states, while it is decaying with time in the noisy case.

More specifically, assuming that the main source of noise is given by the Gaussian uncertainty of the Jaynes-Cummings coupling, we find that for the Fock and (especially) squeezed cases, nonlocality is generated and preserved all along the way (except at some isolated times), although $\langle  {\cal O}_{\rm Bell}\rangle_{\rm max}$ asymptotically approaches 2 (loss of nonlocality). Therefore, these experimental setups are advantageous to test Bell violation. In contrast, for coherent states $\langle  {\cal O}_{\rm Bell}\rangle_{\rm max}$ drops quickly below 2. Concerning entanglement, the asymptotic temporal behaviour is somehow opposite. Starting with a Fock or a squeezed separable (pure) state, the asymptotic mixed state is separable (though for intermediate times is entangled and nonlocal). However, starting with 
any separable coherent (pure) state, the asymptotic mixed state is entangled and satisfies the Peres-Horodecki sufficiency criterium.

We derive also some general relations for pure states of qubit-qudit systems, namely a compact formula for the maximal Bell violation and a general relationship between the latter and the von Neumann entropy of entanglement.

\vspace{0.7cm}
\noindent

\section*{Acknowledgements}

	The authors acknowledge the support of the Spanish Agencia Estatal de Investigacion through the grants ``IFT Centro de Excelencia Severo Ochoa CEX2020-001007-S" and PID2022-142545NB-C22 funded by MCIN/AEI/10.13039/501100011033 and by ERDF. The work of A.B. is supported through the FPI grant PRE2020-095867 funded by MCIN/AEI/10.13039/501100011033.

\newpage
\appendix
\section*{Appendix A: Maximal Bell violation and relation to entropy for qubit-qudit pure states}

Let us first prove the generic formula (\ref{exact}) for $\langle {\cal O}_{\rm Bell} \rangle_{\rm max}$ for pure qubit-qudit states. 

\vspace{0.2cm}
\noindent
Consider a  generic pure state in  the $ {\cal H}_2 \times {\cal H}_d $ Hilbert space 
\begin{equation}
\ket{\psi}   =\ket{0}\otimes \ket{\varphi_0} + \ket{1} \otimes \ket{\varphi_1} \; ,
\end{equation}
where $\{\ket{0}, \ket{1}\}$ is an orthonormal basis of ${\cal H}_2$, but 
$\ket{\varphi_0},\ket{\varphi_1}$ are not necessarily orthogonal or normalized.
The corresponding value of the optimal Bell operator  has to be invariant under local unitary transformations, $U_2\otimes U_d$.  From $U_d$ invariance, it can only depend on   
$(\braket{\varphi_0} , 
 \braket{\varphi_0}{\varphi_1}, 
 \braket{\varphi_1}{\varphi_0}, 
 \braket{\varphi_1})$, which are nothing but the entries of the reduced matrix 
\begin{equation}
\rho_A =  
    \begin{bmatrix}
         \braket{\varphi_0} &   \braket{\varphi_1}{\varphi_0}\\
        \braket{\varphi_0}{\varphi_1} &    \braket{\varphi_1} \\
     \end{bmatrix}.
\end{equation}     
On the other hand, the state can be brought by a $(U_2, U_d)$ transformation to its Schmidt's form
\begin{equation}
\ket{\psi}   =    \cos \theta  \ket{0}_A \otimes  \ket{0}_B  +   \sin \theta   \ket{1}_A \otimes  \ket{1}_B,
\label{Schmidt}
\end{equation}
for some orthonormal $\ket{0}_{A,B}, \ket{1}_{A,B}$, In this basis, $\ket{\varphi_0}=\cos \theta  \ket{0}_B$, $\ket{\varphi_1}=\cos \theta  \ket{1}_B$. Now we can use
the usual $ {\cal H}_2 \times {\cal H}_2 $  result  \cite{POPESCU1992293}
\begin{equation}
\langle {\cal O}_{\rm Bell} \rangle_{\rm max} = 2 \sqrt{1+ 4 \det \rho_A  } \end{equation}
or, in terms of $\ket{\varphi_i}$,
 \begin{equation}
\langle {\cal O}_{\rm Bell} \rangle_{\rm max} = 2 \sqrt{1+ 4\left(\braket{\varphi_0} \braket{\varphi_1} - |\braket{\varphi_0}{\varphi_1}|^2\right)} \; ,
\end{equation}
which coincides with Eq.~(\ref{exact}).

\begin{figure}[h]
\begin{center}
\includegraphics[width=5.5cm,clip=]{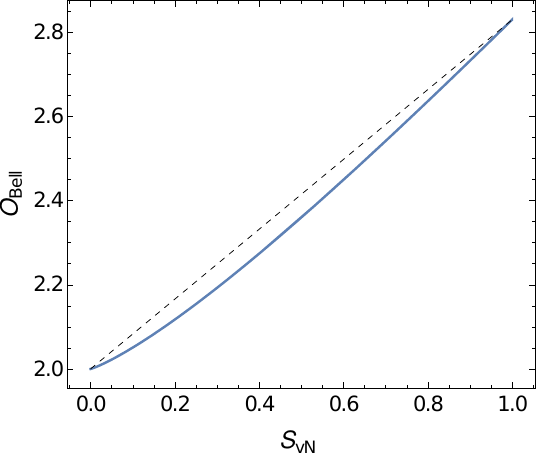}\\
\caption{$\langle {\cal O}_{\rm Bell} \rangle_{\rm max}$ as a function of the von Neumann entropy for a pure qubit-qudit state,
ranging from a separable state  (lower extreme) to a Bell state (upper one). The dashed line is the linear interpolation between the two extrema (exact proportionality).}
\label{fig:linear}
\end{center}
\end{figure}

\vspace{0.3cm}
\noindent
On the other hand, since pure states in $ {\cal H}_2 \times {\cal H}_d $ can be mapped by Schmidt into pures state with $d=2$, Eq.~(\ref{Schmidt}),
entanglement measures for pure $ {\cal H}_2 \times {\cal H}_2 $ states, quantities such as $\langle {\cal O}_{\rm Bell} \rangle_{\rm max} -2$ or the von Neumann (entanglement) entropy  are functions of just the angle $\theta$, namely
\begin{equation}
\begin{array}{lcl}
\langle {\cal O}_{\rm Bell} \rangle_{\rm max} 
& = &  2\sqrt{  1 +  \sin^2  2 \theta},
\\
S_{\text vN} & = &   -  \cos^2{\theta} \log_2 \cos^2{\theta} -  \sin^2{\theta} \log_2 \sin^2{\theta}.
\end{array}           
\end{equation}
This provides a relationship between these two quantities, which is valid for qubit-qudit states. As it is shown in Fig.~\ref{fig:linear}, both quantities are almost proportional.

\newpage

\section*{Appendix B: Asymptotic behavior of separable coherent states}

The partial transposed of the asymptotic density matrix, Eq.\eqref{DensMatInfty},  with respect to the EM-field Hilbert-space indices reads
\be
\rho^{\mathrm{T}_B}(t\to\infty)=\ketbra{e}\otimes \rho_{ee}+\ketbra{e}{g}\otimes \rho_{eg}^{\mathrm{T}}+\ketbra{g}{e}\otimes \rho_{eg}+\ketbra{g}\otimes \rho_{gg}.
\label{TranspInfty}
\ee
Let us consider a state of the form $\ket{\phi_n}=A\,\ket{g}\ket{n}+B\,\ket{e}\ket{n+1}$ with $|A|^2+|B|^2=1$, the product $\rho^{\mathrm{T}_B}(t\to\infty)\,\ket{\phi_n}$ is given by 
\bea
\rho^{\mathrm{T}_B}(t\to\infty)\,\ket{\phi_n}&=&\left[\frac{A}{2}\left(C_g^2\,C_0^2\,\delta_{0,\,n}+C_g^2\,C_n^2+C_e^2\,C_{n-1}^2\right)+B\left(C_g\,C_e\,C_n\,C_{n+1}\right)\right]\ket{g}\ket{n}\\ \ecart
&&+\left[A\left(C_g\,C_e\,C_n\,C_{n+1}\right)+\frac{B}{2}\left(C_e^2\,C_{n+1}^2+C_g^2\,C_{n+2}^2\right)\right]\ket{e}\ket{n+1}.
\eea
The state $\ket{\phi_n}$ is an eigenvector iff $\rho^{\mathrm{T}_B}(t\to\infty)\,\ket{\phi_n}=\mu\,\ket{\phi_n}$, where $\mu$ is the associated eigenvalue. For simplicity, let us rewrite $\rho^{\mathrm{T}_B}(t\to\infty)\,\ket{\phi_n}$ as
\be
\rho^{\mathrm{T}_B}(t\to\infty)\,\ket{\phi_n}=\left[A\,a^n_{11}+B\,a^n_{12}\right]\ket{g}\ket{n}+\left[A\,a^n_{21}+B\,a^n_{22}\right]\ket{e}\ket{n+1},
\ee
where we have introduced the matrix notation:
\bea
a^n_{11}&=&\frac{1}{2}\left(C_g^2\,C_0^2\,\delta_{0,\,n}+C_g^2\,C_n^2+C_e^2\,C_{n-1}^2\right),\\ \ecart
a^n_{12}&=&a^n_{21}=C_g\,C_e\,C_n\,C_{n+1}, \\ \ecart
a^n_{22}&=&\frac{1}{2}\left(C_e^2\,C_{n+1}^2+C_g^2\,C_{n+2}^2\right).
\eea
Solving for $A$, $B$ and $\mu$, we find two conjugate eigenvalues for each $n$: 
\bea
\mu^n_{\pm}=\frac{1}{2}\left(\Tr[a^n]\pm\sqrt{\Tr[a^n]^2-4\det[a^n]}\right),
\eea
with $[a^n]_{ij}=a^n_{ij}$. Obviously, one of these eigenvalues will be negative as long as $\det[a^n]<0$. For fixed values of the coefficients $C_g$ and $C_e$, this condition for $n=0$ is equivalent to $-2 C_e^2+C_g^2\,\alpha^2<0$, i.e.
\be
\alpha^2<\frac{2C_e^2}{C_g^2}
.
\label{AlphCond1}
\ee
Hence, whenever $\alpha^2<\frac{2C_e^2}{C_g^2}$ the eigenvalue $\mu^0_{-}<0$ and therefore $\rho(t\to\infty)$ is entangled. 

For $n\neq0$, the condition $\det[a^n]<0$ translates into
\be
P(n)\equiv C_e^4\, n (2 + n) - 2 C_e^2\, C_g^2\,\alpha^2 (3 + n) + C_g^4\, \alpha^4<0.
\label{AlphCondn}
\ee
In order to get a negative eigenvalue, 
this condition has to be satisfied by at least one integer value of $n$, say $n_{\text{int}}$. In addition, 
$n_{\text{int}}\geq1$ since the case $n=0$ has been already discussed. 

As we have highlighted, the left-hand side of Eq.\eqref{AlphCondn} is a convex quadratic polynomial in $n$, then ensuring that there exists an $n$ for which $P(n)<0$ is equivalent to 
requiring that
%ask whether 
$P(n)$ is negative at its minimum, $P(n_{\min})<0$. Nevertheless, $n_{\min}$ might not be an integer; thus, to check that there exists an integer value $n_{\text{int}}$ leading to $P(n_{\text{int}})<0$, we evaluate both $P(n_{\min})$ and  $P(n_{\min}+1)$. If $P(n_{\min})$ and  $P(n_{\min}+1)$ are both negative we can state that there exists an integer mode $n_{\text{int}}\in [n_{\min},n_{\min}+1]$ leading to $\mu^{n_{\text{int}}}_{-}<0$. 

After a straightforward computation, %Computing the extremal point, we get that 
the minimum of $P(n)$ is attained at $n_{\min}=-1+\frac{C_g^2\,\alpha^2 }{C_e^2}$, hence
\be
P(n_{\min})=-C_e^2(C_e^2+4C_g^2\,\alpha^2),\quad P(n_{\min}+1)=-4C_e^2\,C_g^2\,\alpha^2,
\ee
which are both always negative. To further ensure that $n_{\text{int}}\geq1$ we demand $n_{\min}\geq1$, which is equivalent to 
\be
\alpha^2\geq\frac{2C_e^2}{C_g^2}.
\label{AlphCond2}
\ee
In conclusion,  we have seen that when condition \eqref{AlphCond1} is satisfied, then $\mu^0_{-}<0$ and if \eqref{AlphCond2} does, then $\mu^{n_{\text{int}}}_{-}<0$ . Since both conditions \eqref{AlphCond1} and \eqref{AlphCond2} cover all possible values of $\alpha$, we have proven that given any initial separable coherent state, the asymptotic  $\rho(t\to\infty)$ state is entangled via the Peres-Horodecki criterium.

\bibliographystyle{style2.bst}    % (uses file ``style2.bst"  Includes doi, etc)
\bibliography{references}	 % references a file(s) called  *.bib -DO NOT INCLUDE EXTENSION - 

\providecommand{\href}[2]{#2}\begingroup\raggedright\begin{thebibliography}{10}

\bibitem{Giustina:2015yza}
M.~Giustina et~al., \emph{{Significant-Loophole-Free Test of Bell\textquoteright{}s Theorem with Entangled Photons}}, \href{http://dx.doi.org/10.1103/PhysRevLett.115.250401}{\emph{Phys. Rev. Lett.} {\bf 115} (2015) 250401}, [\href{http://arxiv.org/abs/1511.03190}{{\tt 1511.03190}}].

\bibitem{Hensen:2015ccp}
B.~Hensen et~al., \emph{{Loophole-free Bell inequality violation using electron spins separated by 1.3 kilometres}}, \href{http://dx.doi.org/10.1038/nature15759}{\emph{Nature} {\bf 526} (2015) 682--686}, [\href{http://arxiv.org/abs/1508.05949}{{\tt 1508.05949}}].

\bibitem{Storz:2023jjx}
S.~Storz et~al., \emph{{Loophole-free Bell inequality violation with superconducting circuits}}, \href{http://dx.doi.org/10.1038/s41586-023-05885-0}{\emph{Nature} {\bf 617} (2023) 265--270}.

\bibitem{Zhang2022687}
W.~Zhang et~al., \emph{{A device-independent quantum key distribution system for distant users}}, \href{http://dx.doi.org/10.1038/s41586-022-04891-y}{\emph{Nature} {\bf 607} (2022) 687 – 691}.

\bibitem{Nadlinger:2021uqf}
D.~P. Nadlinger et~al., \emph{{Experimental quantum key distribution certified by Bell's theorem}}, \href{http://dx.doi.org/10.1038/s41586-022-04941-5}{\emph{Nature} {\bf 607} (2022) 682--686}, [\href{http://arxiv.org/abs/2109.14600}{{\tt 2109.14600}}].

\bibitem{Supic20201}
I.~Šupić and J.~Bowles, \emph{{Self-testing of quantum systems: A review}}, \href{http://dx.doi.org/10.22331/Q-2020-09-30-337}{\emph{Quantum} {\bf 4} (2020) 1 – 62}.

\bibitem{Brunner:2013est}
N.~Brunner, D.~Cavalcanti, S.~Pironio, V.~Scarani and S.~Wehner, \emph{{Bell nonlocality}}, \href{http://dx.doi.org/10.1103/RevModPhys.86.419}{\emph{Rev. Mod. Phys.} {\bf 86} (2014) 419}, [\href{http://arxiv.org/abs/1303.2849}{{\tt 1303.2849}}].

\bibitem{Chitambar:2018rnj}
E.~Chitambar and G.~Gour, \emph{{Quantum resource theories}}, \href{http://dx.doi.org/10.1103/revmodphys.91.025001}{\emph{Rev. Mod. Phys.} {\bf 91} (2019) 025001}, [\href{http://arxiv.org/abs/1806.06107}{{\tt 1806.06107}}].

\bibitem{PhysRevA.44.6023}
S.~J.~D. Phoenix and P.~L. Knight, \emph{{Establishment of an entangled atom-field state in the Jaynes-Cummings model}}, \href{http://dx.doi.org/10.1103/PhysRevA.44.6023}{\emph{Phys. Rev. A} {\bf 44} (1991) 6023--6029}.

\bibitem{Halder:2023iod}
P.~Halder, R.~Banerjee, S.~Roy and A.~S. De, \emph{{Hybrid nonlocality via atom photon interactions with and without impurities}},  \href{http://arxiv.org/abs/2302.11513}{{\tt 2302.11513}}.

\bibitem{Blais:2020oqf}
A.~Blais, S.~M. Girvin and W.~D. Oliver, \emph{{Quantum information processing and quantum optics with circuit quantum electrodynamics}}, \href{http://dx.doi.org/10.1038/s41567-020-0806-z}{\emph{Nature Phys.} {\bf 16} (2020) 247--256}.

\bibitem{Jaynes:1963zz}
E.~T. Jaynes and F.~W. Cummings, \emph{{Comparison of quantum and semiclassical radiation theories with application to the beam maser}}, \href{http://dx.doi.org/10.1109/PROC.1963.1664}{\emph{IEEE Proc.} {\bf 51} (1963) 89--109}.

\bibitem{Shore19931195}
B.~W. Shore and P.~L. Knight, \emph{{The Jaynes–Cummings model}}, \href{http://dx.doi.org/10.1080/09500349314551321}{\emph{Journal of Modern Optics} {\bf 40} (1993) 1195 – 1238}.

\bibitem{Larson:2022jvs}
J.~Larson and T.~Mavrogordatos, \emph{{The Jaynes–Cummings Model and Its Descendants}}, \href{http://dx.doi.org/10.1088/978-0-7503-3447-1}{\emph{IOP Publishing} (2021) }, [\href{http://arxiv.org/abs/2202.00330}{{\tt 2202.00330}}].

\bibitem{Clauser:1969ny}
J.~F. Clauser, M.~A. Horne, A.~Shimony and R.~A. Holt, \emph{{Proposed experiment to test local hidden variable theories}}, \href{http://dx.doi.org/10.1103/PhysRevLett.23.880}{\emph{Phys. Rev. Lett.} {\bf 23} (1969) 880--884}.

\bibitem{PhysRevLett.88.040406}
Z.-B. Chen, J.-W. Pan, G.~Hou and Y.-D. Zhang, \emph{{Maximal Violation of Bell's Inequalities for Continuous Variable Systems}}, \href{http://dx.doi.org/10.1103/PhysRevLett.88.040406}{\emph{Phys. Rev. Lett.} {\bf 88} (2002) 040406}.

\bibitem{ZHANG20111765}
B.~Zhang, Z.-R. Zhong and S.-B. Zheng, \emph{{Quantum nonlocality for two-mode correlated states based on generalized pseudospin operators}}, \href{http://dx.doi.org/https://doi.org/10.1016/j.physleta.2011.03.029}{\emph{Phys. Lett. A} {\bf 375} (2011) 1765--1768}.

\bibitem{PhysRevLett.98.013601}
Y.~Wu and X.~Yang, \emph{{Strong-Coupling Theory of Periodically Driven Two-Level Systems}}, \href{http://dx.doi.org/10.1103/PhysRevLett.98.013601}{\emph{Phys. Rev. Lett.} {\bf 98} (2007) 013601}.

\bibitem{PHOENIX1988381}
S.~Phoenix and P.~Knight, \emph{{Fluctuations and entropy in models of quantum optical resonance}}, \href{http://dx.doi.org/https://doi.org/10.1016/0003-4916(88)90006-1}{\emph{Annals of Physics} {\bf 186} (1988) 381--407}.

\bibitem{gea1992new}
J.~Gea-Banacloche, \emph{{A new look at the Jaynes-Cummings model for large fields: Bloch sphere evolution and detuning effects}}, \href{http://dx.doi.org/10.1016/0030-4018(92)90082-3}{\emph{Optics Communications} {\bf 88} (1992) 531--550}.

\bibitem{Capasso:1973wt}
V.~Capasso, D.~Fortunato and F.~Selleri, \emph{{Sensitive observables of quantum mechanics}}, \href{http://dx.doi.org/10.1007/BF00669912}{\emph{Int. J. Theor. Phys.} {\bf 7} (1973) 319--326}.

\bibitem{Gisin:1991vpb}
N.~Gisin, \emph{{Bell's inequality holds for all non-product states}}, \href{http://dx.doi.org/10.1016/0375-9601(91)90805-I}{\emph{Phys. Lett. A} {\bf 154} (1991) 201--202}.

\bibitem{Collins:2002sun}
D.~Collins, N.~Gisin, N.~Linden, S.~Massar and S.~Popescu, \emph{{Bell Inequalities for Arbitrarily High-Dimensional Systems}}, \href{http://dx.doi.org/10.1103/PhysRevLett.88.040404}{\emph{Phys. Rev. Lett.} {\bf 88} (2002) 040404}.

\bibitem{Pironio_2014}
S.~Pironio, \emph{{All Clauser--Horne--Shimony--Holt polytopes}}, \href{http://dx.doi.org/10.1088/1751-8113/47/42/424020}{\emph{Journal of Physics A: Mathematical and Theoretical} {\bf 47} (2014) 424020}.

\bibitem{HORODECKI1995340}
R.~Horodecki, P.~Horodecki and M.~Horodecki, \emph{{Violating Bell inequality by mixed spin-1/2 states: necessary and sufficient condition}}, \href{http://dx.doi.org/https://doi.org/10.1016/0375-9601(95)00214-N}{\emph{Physics Letters A} {\bf 200} (1995) 340--344}.

\bibitem{Bernal:2024fio}
A.~Bernal, J.~A. Casas and J.~M. Moreno, \emph{{Optimal Bell inequalities for qubit-qudit systems}},  \href{http://arxiv.org/abs/2404.02092}{{\tt 2404.02092}}.

\bibitem{Peres:1996dw}
A.~Peres, \emph{{Separability criterion for density matrices}}, \href{http://dx.doi.org/10.1103/PhysRevLett.77.1413}{\emph{Phys. Rev. Lett.} {\bf 77} (1996) 1413--1415}, [\href{http://arxiv.org/abs/quant-ph/9604005}{{\tt quant-ph/9604005}}].

\bibitem{Horodecki:1996nc}
M.~Horodecki, P.~Horodecki and R.~Horodecki, \emph{{On the necessary and sufficient conditions for separability of mixed quantum states}}, \href{http://dx.doi.org/10.1016/S0375-9601(96)00706-2}{\emph{Phys. Lett. A} {\bf 223} (1996) 1}, [\href{http://arxiv.org/abs/quant-ph/9605038}{{\tt quant-ph/9605038}}].

\bibitem{POPESCU1992293}
S.~Popescu and D.~Rohrlich, \emph{Generic quantum nonlocality}, \href{http://dx.doi.org/https://doi.org/10.1016/0375-9601(92)90711-T}{\emph{Physics Letters A} {\bf 166} (1992) 293--297}.

\end{thebibliography}\endgroup

\end{document}